# Stain-Induced Band Inversion and Topological Nontriviality in Antimonene

Mingwen Zhao

*School of Physics and State Key Laboratory of Crystal Materials, Shandong University, Jinan 250100, Shandong, China*

Antimonen is a novel two-dimensional (2D) semiconducting material of group V elements proposed in a recent literature [Zhang et al., *Angew. Chem. Int. Ed.* 54, 1-5 (2015)]. Using first-principles calculations, we demonstrated that the buckled configuration of antimonene enables it sustain large tensile strain up to 20%. Band inversion takes place in the vicinity of the Γ point as the tensile strain is larger than 14.5%, leading to six tilted Dirac cones in the Brillouin zone. Spin-orbital coupling (SOC) effect opens up a topologically nontrivial bulk band gap at the Dirac points, exhibiting the features of 2D topological insulators characterized by a nonzero $Z_2$ topological invariant. The tunable bulk band gap, 101-560 meV, make the antimonene a promising candidate material for achieving quantum spin Hall effect (QSH) at high temperatures which meet the requirement of future electronic devices with low power consumption.

E-mail: zmw@sdu.edu.cn

Motivated by the successful discovery and application of graphene, there has been extensive research into other two-dimensional (2D) materials. Some of them are superior to graphene in specific aspects. For example, graphene was firstly proposed as a 2D topological insulator with a bulk band gap due to spin-orbital coupling (SOC) and gapless edge states protected by time-reversal symmetry [1-4]. The edge states characterized by Dirac-cone-like linear energy dispersion are quite promising for the realization of conducting channels without dissipation due to the robustness against backscattering [5-6]. However, the SOC effect in graphene is very week SOC, leading to an unobservably small bulk band gap ($\sim 10^{-3}$ meV). The critical temperature to achieve dissipationless transport in graphene is unrealistically low [7-10]. This drawback can be overcome in other 2D materials of group IV elements, such as silicene [11], germanene [11], and stanene [12], and those of group III-V compounds, such as GaAs [13] and GaBi [14]. The SOC in these materials is much stronger than that in graphene and the bulk band gaps are consequently larger. For example, upon fluorination, the nontrivial bulk band gap in GaB monolayer can be as large as 0.95 eV. Therefore, the critical temperature for achieving dissipationless transport in this material will be above room temperature. Such significant improvement is related to the unique electron band inversion and buckled configurations of these 2D materials.

The 2D materials of group V elements are drawing considerable interest because of the unique properties. Phosphorene, a 2D material that can be isolated through mechanical exfoliation from layered black phosphorus, is a normal semiconductor with a sizable band gap (1.2-2.2 eV). A field effect transistor (FET) action has been demonstrated in a few layer phosphorene by manipulating the doping level via back-gate voltage, leading to on-off ratio of the order of $10^5$ [15] . A recent theoretical work shows that few-layer phosphorene can be converted to a 2D topological insulator with a bulk band gap of about 5 meV by applying perpendicular electric field [16]. The small bulk band gap is due to the weak SOC of light elements. For bismuth which is the heaviest group V element, SOC is naturally strong. Therefore, Bi(111) bilayer was predicted to be a 2D topological insulator with a large bulk band gap [17]. This 2D bismuth material has been successfully grown on $Bi_2Te_3$ substrates[18] or locally exfoliated from bulk crystal[19]. The existence of the edge states in Bi(111) bilayer has been reported[20,21]. Very recently, antimonene, a new 2D material of group V element, was proposed on the basis of first-principles calculations [22]. The isolation of antimonene was expected to be achieved by exfoliating layered Sb crystal, thanks to the weak interlayer interaction. Although bulk Sb crystal is semimetallic, antimonene becomes semiconducting when it is chinned to one atomic layer. However, unlike Bi(111) bilayer, antimonene proposed in that work is only a normal semiconductor with a band gap of 2.28 eV [22]. If the antimonene can be tuned to a nontrivial topological insulator becomes an interesting issue.

Using first-principles calculations within density of functional theory (DFT), we demonstrated that this goal can be reached by applying a biaxial tensile strain larger than 14.5%. The transition from a normal semiconductor to a nontrivial topological insulator is attributed to the strain-induced band inversion in the vicinity of the Γ point. The bulk band gap due to SOC increases with the increase of tensile strain. The buckled configuration of antimonene enables it to endure large tensile strain up to 20% and the bulk band gap can be as large as 560 meV. This interesting result implies that antimonene is a promising candidate material for achieving quantum spin Hall effect (QSH) at high temperatures which meet the requirement of future electronic devices with low power consumption.

All the calculations were performed using the plane wave basis Vienna ab initio simulation package known as VASP code [23-24]. The ion-electron interactions are treated using projector-augmented-wave potentials [25]. The electron exchange-correlation functional was treated using generalized gradient approximation (GGA) in the form proposed by Perdew, Burke, and Ernzerhof (PBE) [26]. The atomic positions were relaxed until the maximum force on each atom was less than 0.01 eV/Å. The energy cutoff of the plane waves was set to 600 eV with the energy precision of $10^{-8}$ eV. The Brillouin zone (BZ) was sampled by using an 11×11×1 Gamma-centered Monkhorst-Pack grid. Antimonene is modeled by unit cells repeated periodically on the x-y plane, while a vacuum region of about 15 Å is applied along the z-direction to exclude the interactions between images. In the electronic structure calculations, an accurate Heyd-Scuseria-Ernzerhof (HSE) screened Coulomb hybrid density functional [27] was adopted to correct the shortcoming of the PBE functional which always underestimates the band gaps. SOC was included by a second variational procedure on a fully self-consistent basis.

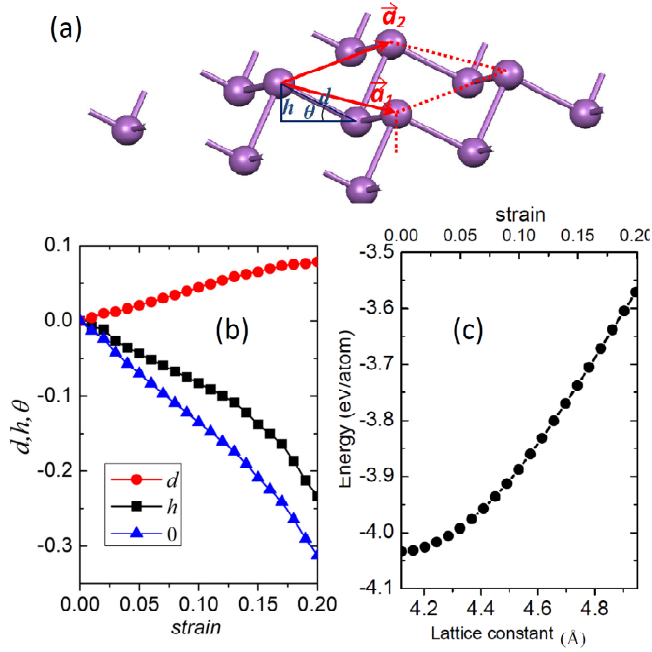

**Figure 1.** (a) Schematic representation of the buckled configuration of antimonene. (b) structural evolution represented by three parameters indicated in (a) of antimonene under tensile strain. (c) Variation of energy in response to tensile strain.

The antimonene has the point group symmetry of $D_{3d}$ with spatial inversion included, as shown in Fig. 1(a). Similar to the case of Bi(111) bilayer, it exhibits a bipartite honeycomb lattice with A and B sublattices. The two sublattices have different heights, forming a buckled configuration. At the equilibrium state, the Sb-Sb bond length is about 2.89 Å and the height of buckling (h) is 1.64 Å with an angle of $\theta = 26.8°$, as indicated in Fig. 1(a). The length of the basis vectors is 4.12 Å. slightly longer than that reported in Ref [22]. The binding energy calculated from the energy difference between antimonene and isolated Sb atoms is about -4.034 eV/atom.

Buckled configuration is always expected to sustain a larger tensile strain than planner one. We therefore studied the structural and energetic evolution under biaxial tensile strain. In our calculations, the biaxial tensile strain was applied by fixing the lattice constant to a series of values longer than that of the equilibrium state and optimizing the atomic coordinates. The variation of Sb-Sb distance ($d$), buckling height ($h$) and buckling angle ($\theta$) as a function of tensile strain ($\tau$) is plotted in Fig. 1(b). The binding energy of the antimonene at each tensile strain is shown in Fig. 1(c). With the increase of tensile strain, the Sb-Sb distance increases gradually, while both buckling height and buckling angle decrease rapidly. At $\tau = 0.2$, the Sb-Sb bond is only stretched by 7.9%, but $h$ and $\theta$ are reduced by 23.4% and 32.4%, respectively. The Sb-Sb covalent bonds of antimonene are preserved in this tensile strain range. This leads to slight energy increase, as shown in Fig. 1(c). The energy increase at $\tau = 0.2$ is about 0.463 eV/atom, only 11.5% of the value at the equilibrium state.

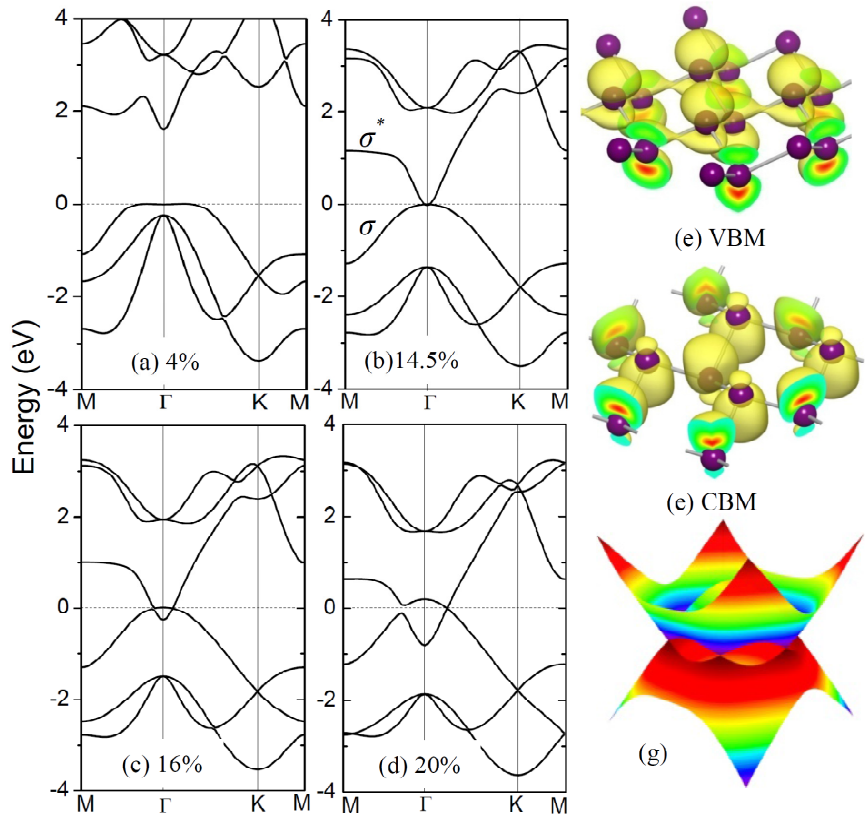

**Figure 2.** (a)-(d) Electronic band structures of antimonene under different tensile strains obtained from DFT calculation within HSE functional. The energy at the Fermi level was set to zero. (e),(f) Electron density profiles of the wavefuction of the valence band maximum (VBM) and conduction band minimum (CBM) of the antimonene under the tensile strain of 4%. (g) Two-dimensional electron band structure of the antimonene under the tensile strain of 20% in the vicinity of the Dirac points.

The electronic structure modification of the antimonone in response to the tensile strain is shown in Fig. 2. At the tensile strain of τ = 0.04, antimonene converts to a direct-band-gap semiconductor with a band gap of 1.98 eV at the Γ point. The density profile of the electron wavefunction at valence band maximum (VBM) and conduction band minimum (CBM) can be featured as σ and σ* states of $p_z$ atomic orbitals of Sb, as indicated by the isosurfaces of wavefunctions shown in Fig. 2(e) and 3(g), respectively. The band gap decreases with the increase of tensile strain and comes to close at τ = 0.145, as shown in Fig. 2(b). When the tensile strain is further increased, σ-σ* band inversion takes place in the region near Γ, as shown Fig. 2(c). A small band gap along the Γ-M direction is opened up due to the σ-σ* coupling, while the meeting point between valence and conduction bands is preserved along the Γ-K direction, giving rise to six Dirac cones with six-fold symmetry in BZ, as shown in Fig.2 (f). It is noteworthy that the Dirac cone in this stretched antimonene is titled with strong directional anisotropy in the reciprocal space, which would lead to interesting properties that differ significantly from the Dirac-Fermions in graphene.

To illustrate the band inversion mechanism explicitly, we propose a tight-binding model of $s, p_x, p_y$ and $p_z$ atomic orbitals. The effective Hamiltonian is taken as:

$$H_{TB} = \sum_{i,\alpha} \varepsilon_i^\alpha c_i^{\alpha+} c_i^\alpha + \sum_{<i,j>,\alpha,\beta} t_{ij}^{\alpha\beta} (c_i^{\alpha+} c_j^\beta + h.c.)$$

Here, $\varepsilon_i^\alpha$, $c_i^{\alpha+}$, and $c_i^\alpha$ represent the on-site energy, creation, and annihilation operators of an electron at the α-orbital of the *i*-th atom. The $t_{ij}^{\alpha\beta}$ parameter is the nearest-neighbor hopping energy of an electron between an α-orbital of *i*-th atom and β-orbital of *j*-th atom, $\alpha, \beta \in (s, p_x, p_y, p_z)$. According to TB theory, the hopping energies can be evaluated on the basis of four parameters ($V_{ss\sigma}$, $V_{sp\sigma}$, $V_{pp\sigma}$, and $V_{pp\pi}$) in combined with the atomic coordinates of the antimonene (see Supporting Materials). For simplification, the values of $V_{ss\sigma}$, $V_{sp\sigma}$, $V_{pp\sigma}$, and $V_{pp\pi}$ were supposed to be independent of the tensile strain. This is reasonable because the Sb-Sb bond length changes slightly, especially at small tensile strain. We started from the semiconducting antimonene with a direct band gap which can be reproduced by the TB Hamiltonian.

When a tensile strain is applied to the antimonene, the buckling angle was changed to the data of DFT calculations to describe the structure deformation in response to the tensile strain. Using this simple TB Hamiltonian, we found that the band gap decreases with the decrease of buckling angle and comes to close at θ = 27.7°, corresponding to a tensile strain of 14.5%, which is in good agreement with DFT results. As the buckling angle is further reduced, band inversion takes place in the vicinity of Γ point. Interestingly, the band gap opened along the Γ-M direction due to the σ-σ* coupling can also be reproduced using this TB model. This implies that the strain-induced band inversion is mainly due to reducing buckling angle under tensile strain, rather than the changes of Sb-Sb bond. We also varied the $V_{ss\sigma}$, $V_{sp\sigma}$, $V_{pp\sigma}$, and $V_{pp\pi}$ to reflect the elongation of the Sb-Sb bond, but found that the evolution of the band lines is insensitive to changes of these parameters. The strain-induced band inversion in the antimonene differs significantly from the band inversion in Bi bilayer which is caused by spin-orbital coupling [28].

We then turned on SOC in the band structure calculations. It is not surprising that a band gap is opened at the Dirac point of the antimonene, as shown in Fig. 3(a). More interestingly, the SOC gap increases with the increase of tensile strain and can be as large as 560 meV at the tensile of τ = 0.2, as shown in Fig. 3(c). The variation trend of the SOC gap can be well reproduced by introducing a spin-orbital component into the TB Hamiltonian(see Fig. 3(b),(c) and the Supporting Information). The deviation of the SOC gap between DFT and TB model at large tensile strain is related to the changes of $V_{ss\sigma}$, $V_{sp\sigma}$, $V_{pp\sigma}$, and $V_{pp\pi}$ due to the elongation of the Sb-Sb bond which was omitted in our TB model.

The topological nontriviality of the stretched antimonene can be confirmed by calculating the topological invariant $Z_2$. For a lattice with inversion symmetry, the $Z_2$ index can be deduced from the knowledge of the parities of the four time-reversal and parity invariant points at BZ, without having to know about the global properties of the energy bands. The honeycomb lattice of the antimonene has the four time-reversal invariant momenta at the point of $\Gamma_i = n_1 \vec{b_1} + n_2 \vec{b_2}$, with $\vec{b_1}$ and $\vec{b_2}$ being the base vectors of the reciprocal lattice and $n_1, n_2 \in \{0, 1/2\}$. The $Z_2$ invariant $\nu$ is defined by

$$(-1)^\nu = \prod_i \delta_i \text{ with } \delta_i = \prod_{m=1}^{N} \xi_{2m}(\Gamma_i)$$

for 2N occupied bands. $\xi_{2m}(\Gamma_i) = \pm 1$ is the parity eigenvalue of the *2m*-th occupied energy band at the time-reversal invariant momentum $\Gamma_i$. Our first-principles calculations showed $\delta_i$ has the values of (+), (-), (-), and (-) at (0, 0), (1/2, 0), (0, 1/2), and (1/2, 1/2) time-reversal momenta when the band inversion takes place as the tensile strain is larger than 14.5%, as shown in Fig. 3(d). The topological invariant is therefore $Z_2=1$, indicating that the stretched antimonene is a topological insulator.

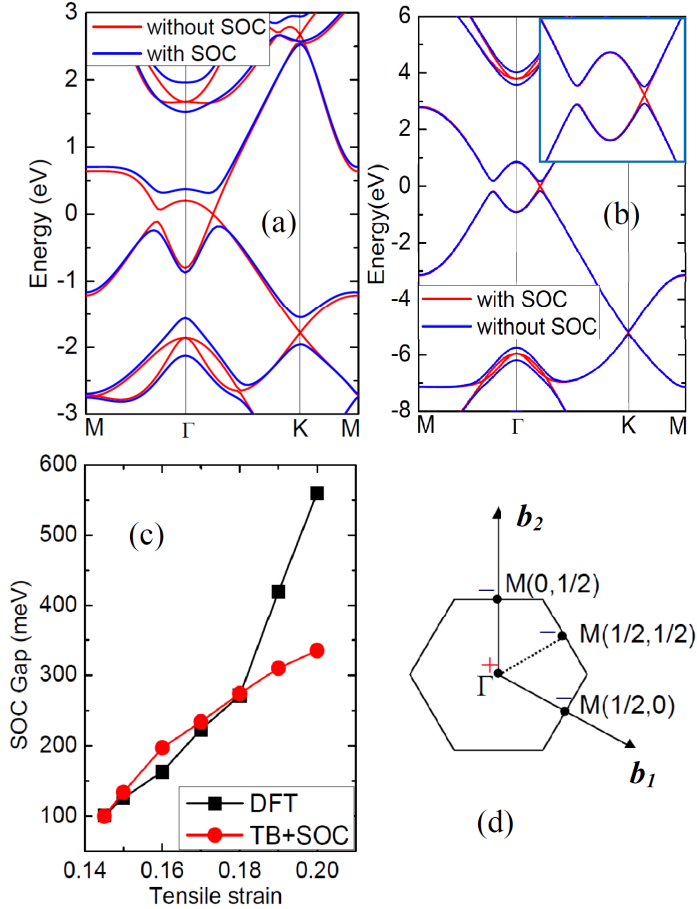

**Figure 3.** (a),(b) Electronic band structures of the antimonene under the tensile strain of 20% obtained from (a) DFT calculations within HSE functional and (b) tight-binding model with (blue lines) and without (red lines) spin-orbital coupling (SOC). (c) Evolution of band gap opened due to SOC in response to tensile strain. (d) The parities of the occupied bands at the time-reversal invariant momentum.

In summary, based on the first-principles calculations combined with a tight-binding model, we demonstrated that the trivial semiconducting antimonene can be tuned to a topological insulator by applying a biaxial tensile strain larger than 14.5%. The electronic structure transition is closely related to the strain-induced σ-σ* band inversion in the vicinity of the Γ point. The SOC gap increases with the increase of tensile strain. The buckled configuration of antimonene enables it sustain large tensile strain up to 20%, which gives rise to a SOC gap of 560 meV. These interesting results make antimonene a promising candidate material for achieving quantum spin Hall effect (QSH) at high temperatures which meet the requirement of future electronic devices with low power consumption.

**ACKNOWLEDGMENT**

This work is supported by the National Basic Research Program of China (No.2012CB932302), the National Natural Science Foundation of China (Nos.91221101, 21433006), the 111 project (No. B13029), the Taishan Scholar Program of Shandong, and the National Super Computing Centre in Jinan.